# Topological and magnetic phase transition in silicene-like zigzag nanoribbons

**Xiao Long Lü, Yang Xie and Hang Xie**[*]

Department of Physics, Chongqing University, Chongqing, 401331, P. R. China

Email: xiehangphy@cqu.edu.cn

**Abstract**

Spin-orbital interactions (SOI) in silicene results in the quantum spin Hall effect, while the Hubbard-induced Coulomb interaction in zigzag nanoribbons often generates a band gap with the anti-ferromagnetic (AF) spin orders on two edges. In this paper we systematically study these two joint contributions to the zigzag silicene-like nanoribbons (zSiNR). Some topological and magnetic phase transitions are investigated with different material parameters and external fields. We find when the ribbon width or the SOI value exceeds some critical value, the SOI may overcome the Coulomb interaction and the system transits from a band insulator to a topological insulator: the quantum-spin-Hall or the spin quantum-anomalous Hall state. We also find some magnetic phase transition exist in the Hubbard-dominated zSiNR systems when the exchange field or the electric field goes beyond some critical values. At last we observe a double topological/magnetic phase transition in a Hubbard-SOI-balanced zSiNR system before the magnetic and topological phases are destroyed by a strong electric field.

## 1. Introduction

Recently, the two-dimensional (2D) materials have attracted a lot of research interests due to their special electronic, mechanical and optical properties [1, 2]. The basis 2D materials: graphene has the linear dispersion relation around the Fermi level. This leads to the Klein tunneling effect which is described by the massless Dirac equation in the low-energy expansion [3, 4].

Beyond graphene, silicene, and similar 2D materials such as germanene and stanene have the buckled structures which results in the relatively large spin-orbital interaction (SOI) [5-7]. This

SOI opens a gap in the bulk systems but generates the gapless edge states in the nanoribbons [8-10]. These edges states are topologically-protected because they are robust against small perturbations. The materials with bulk energy gap but gapless edges states are the topological insulators (TI). There are plenty of the TI behaviors such as the quantum spin Hall (QSH) effect [11, 12], the quantum valley Hall (QVH) effect [13, 14], quantum anomalous Hall (QAH) effect [15, 16]. These states results from the additional Berry phases characterized by the Chern numbers[8].

The edges states in the QSH are spin-dependent which can be used in the spin filter devices. There are a lot of such researches based on the quantum transport theories with the tight-binding (TB) model [17-20]. The external exchange magnetic field or the electric filed is often employed on these zigzag silicene-like nanoribbons (zSiNR) to regulate the spin transport properties. Besides the model systems, there are some first-principles calculations on the zSiNR, which shows the anti-ferromagnetic (AF) or ferromagnetic (FM) edges states in the band structures [21, 22]. These magnetic edge states come from the Coulomb repulsive interactions. Similar edge states in the graphene nanoribbons (GNR) have been deeply studied with the Hubbard model and the first-principles models [23-27]. Our previous study also shows there exist various excited spin density waves (ESDW) in GNR [28]. For the zSiNR, someone uses the Hubbard model to study the magnetic band structures and the thermoelectric properties [29-31].

In these studies of the zSiNR, the TB model without the Hubbard term shows gapless edges states in the band structures [17, 19, 20], but the first-principles or Hubbard-model calculations show a band gap for the ground state (AF state) [21, 22, 30, 31]. This discrepancy is due to different models. However, with the small SOI of silicene (3.9 meV), the bands near the Fermi level of the zSiNR become some flat at the Brillouin zone boundary [5, 11]. This leads to a large density of states (DOS), which results in a strong Coulomb interaction [32]. So the Hubbard term's contribution may not be neglected, especially in the narrow ribbons with very flat bands near the Fermi level. In this situation, reconciling these two models is necessary.

In this paper we give a self-consistent study for the Hubbard model in the zSiNR. We find when increasing the width of zSiNR, or the intensity of SOI, the gap of the energy band (in the AF state) will approach to be closed, which is the case of the TB model without the Hubbard term. With our new-developed method of the Berry curvature calculation, we find in this process the

Chern number is changed. So there exists a topological phase transition when the gap tends to be closed.

In the narrow zSiNR, the Coulomb interaction overcomes the SOI and a gap with magnetism occurs. It is worthwhile to further investigate the possible magnetic phase transition under different conditions with the Hubbard model. Similar phase transitions in zigzag GNRs have been studied in the recent theoretical and experimental works [33-35].With the shift of the chemical potential, the magnetism of the band structures are proposed to be changed [33]. People also measured a metal-insulator transition with different widths of zigzag GNRs [34]. In this transition the AF state changes into the FM state due to the change of self-consisted chemical potential. Furthermore, the magnetism on the GNR edges is found to oscillate when increasing the ribbon width [35].

In this paper we use the Hubbard-TB model to explore some topological and magnetic phase transition in the presence of exchange magnetic field and perpendicular electric field. These external fields are available in the current experiments. Some interesting magnetic phase transitions occur in the zero and the finite temperatures when changing the external fields. We find near these transitions, the energy drops suddenly and some spin bands merge after the critical field. We give a detailed explanation of these transition processes based on the principle of the local minimum of the total energy in the configuration space [28]. With a strong electric field, there exists a paramagnetic state without any magnetism. In the large-SOI systems, we also observe the double topological/magnetic phase transitions in the zSiNRs with different electric fields.

This paper is divided into the following parts: Section II is about the models and the methods in our calculations; Section III shows our calculation results and analysis. The conclusions are given in the Section IV.

## 2. Model and methods

In this paper, we use the tight-binding model to describe the zSiNR systems. The system can be described by the following Hamiltonian [19, 31]

$$H = -t \sum_{<i,j>\alpha} c_{i\alpha}^{\dagger} c_{j\alpha} + i \frac{\lambda_{SO}}{3\sqrt{3}} \sum_{<<i,j>>\alpha\beta} v_{ij} c_{i\alpha}^{\dagger} \sigma_{\alpha\beta}^{z} c_{j\beta} + E_z l \sum_{i\alpha} u_i c_{i\alpha}^{\dagger} c_{i\alpha} + M_{FM} \sum_{i\alpha} c_{i\alpha}^{\dagger} \sigma_z c_{i\alpha}$$
$$+ M_{AF} \sum_{i\alpha} u_i c_{i\alpha}^{\dagger} \sigma_z c_{i\alpha} + U \sum_i (n_{i\uparrow} <n_{i\downarrow}> + n_{i\downarrow} <n_{i\uparrow}> - \frac{1}{2})$$
(1)

where $c_{i\alpha}^{\dagger}$ ($c_{i\alpha}$) is the creation (annihilation) operator for an electron with spin $\alpha = \uparrow, \downarrow$ at the lattice point $i$. The spin polarized axis along the z axis perpendicular to the plane of the nanoribbon. The first term is restricted here to the electron hopping between the nearest-neighboring sites, with the corresponding hopping parameter *t*. For instance, *t*=1.6eV in silicene. The second term refers to the intrinsic SOI following from the electron hopping between the next-nearest-neighbors in the silicene nanoribbon, where $\lambda_{so}$ is the SOI parameter, while $v_{ij}$ =1($v_{ij}$ =-1) when the hopping path from the second-nearest-neighboring sites $i$ to $j$ in the zSiNR is clockwise (anticlockwise) with respect to the positive z-axis. $\sigma_{\alpha\beta}^{z}$ is the z component Pauli matrix with the spin indices $\alpha$ and $\beta$. The third term represents the staggered sub-lattice electric potential (with the electric field $E_z$ and the buckle height *l*), where $u_i = \pm 1$ for the A (B) site. The fourth term represents the FM exchange magnetization. The exchange field $M_{FM}$ may arise due to the proximity coupling to a ferromagnet by depositing Fe atoms to the silicene surface or depositing silicene to a ferromagnetic insulating substrate. The fifth term is the AF exchange field, in which the A (B) sublattice atoms experience difference exchange fields. The sixth term is the Hubbard interaction in the mean-field approximation where $n_{i\alpha}$ are the particle number operators for site $i$ with the spin $\alpha$, $<n_{i\alpha}>$ is the mean value averaged in the first Brillouin zone.

In the Hubbard term, $<n_{i\alpha}>$ is evaluated by the following formula [27]

$$<n_{i\alpha}> = \frac{a}{2\pi} \int_0^{2\pi/a} n_{i\alpha}(k) dk \qquad (2)$$

where $n_{i\alpha}(k)$ is the spin-dependent electron density at site $i$ with the Bloch-wavevector *k*. The parameter *a* denotes the lattice constant of a unit cell of the zSiNR. For the two-dimensional (2D) model for the Berry curvature calculation, $<n_{i\alpha}>$ is evaluated by the formula below

$$<n_{i\alpha}> = \frac{a}{2\pi}\frac{b}{2\pi} \int_0^{2\pi/b} \int_0^{2\pi/a} n_{i\alpha}(k_x, k_y) dk_x dk_y, \qquad (3)$$

where *a* and *b* are the spacing distance of the unit cell in the x and y direction.

The self-consistent (SC) calculation is employed in the Hubbard system for the electron

density $n_{i\alpha}(k)$ [28, 31]

$$n_i^{\uparrow(\downarrow)}(k) = \sum_p^{all} |A_{i,p}^{\uparrow(\downarrow)}(k)|^2 f(E_p(k),\mu), \qquad (4)$$

$$\mathbf{H}(<n^{\downarrow(\uparrow)}>,k)\mathbf{A}_p^{\uparrow(\downarrow)} = E_p(k)\mathbf{A}_p^{\uparrow(\downarrow)}, \qquad (5)$$

where $f(E,\mu) = 1/(\exp((E-\mu)/k_BT)+1)$, is the Fermi-Dirac distribution with the chemical potential $\mu$; 'all' means the summation is utilized for the all orbitals; $A_{i,p}^{\uparrow(\downarrow)}(k)$ is the eigenvector of $i^{th}$ site for the $p^{th}$ eigenvalue ($E_p(k)$) with the Bloch wavevector $k$. We use Eqs. (1)-(4) to solve this SC problem. The chemical potential of a primitive nanoribbon is equal to the corresponding Fermi level ($E_F$) and in our TB model it is set to zero for simplicity.

To investigate the topological properties of these zSiNR, we develop a quasi-1D method for the Berry curvature calculation. This method actually is a 2D model with a weak coupling in the y direction (see Fig.2 (a) later). The Hamiltonian of the 2D model will be given in Sec. IIIA. We see that when the coupling matrix (including the hopping integrals and the SOI) in the y direction are reduced to zero, the supercell system gradually transits from a standard 2D system to a quasi-1D system. The formula for calculating the Berry curvature is given below [15]

$$\Omega_n = \sum_{n \neq n'} \frac{-2\,\mathrm{Im}[<\psi_{nk}|\frac{\partial H}{\partial k_x}|\psi_{n'k}><\psi_{n'k}|\frac{\partial H}{\partial k_y}|\psi_{nk}>]}{(E_n - E_{n'})^2}. \qquad (6)$$

The corresponding Chen number is

$$c_n = \frac{1}{2\pi}\iint \Omega_n d^2k. \qquad (7)$$

where the integral is calculated in the first Brillouin zone. With calculating the Berry curvature and Chen numbers by the formula above, we may investigate the topological properties of these Hubbard-involved zSiNR systems.

### 3. Results and Discussions

**3.1 Topological phase transition in zSiNR with different widths and SOI**

Firstly we use this TB+Hubbard model to investigate the band structure and energy changes of zSiNR with different ribbon widths and SOI. The hopping integral ($t$) is set as 1.6 eV and the

Hubbard value (U) is set as 1.4 eV for the zigzag silicene ribbon [29] in this paper. The SOI value ($\lambda_{SO}$) for silicene is 3.9 meV, while $\lambda_{SO}$ for germanene and stanene is 0.043 eV or $0.033t$ ($t$=1.3 eV) and 0.1 eV or $0.077t$ ($t$=1.3 eV) respectively[36]. For some fluorinated zSiNR, $\lambda_{SO}$ may be even larger [36]. In order to observe the topological phase transition, the small $\lambda_{SO}$ (3.9 meV) needs a very large ribbon width with heavy calculations. So in this part we choose a larger SOI value: $\lambda_{SO}$ =0.05 eV=$0.031t$, which is very close to the germanene.

In the SC band calculations with the Hubbard term, we considerate the AF configuration which is the ground state [22]. The band results of different ribbon widths are shown in Fig. 1(a)-(d). One unit cell of the zSiNR consists of $N_y$ 4-atom units (see Fig. 2(a)). Here we only consider the zero temperature case. We see all the bands are spin-degenerate. In the narrow zSiNR ($N_y$=5), the two edge bands near the Fermi level (the highest occupied orbital (HOMO) and the lowest occupied orbital (LUMO)) have an gap, which is similar to the AF state of GNR [27]. The asymmetry of the gap comes from the SOI. When increasing the ribbon width, the gap becomes decreased until the HOMO and LUMO bands tend to contact (Fig. 1 (b)). In a very wide ribbon, the two edge bands almost contact (Fig. 1(c)), which is very similar to the band structure of the TB model without the Hubbard term [8, 20]. Actually there exists a very tiny gap in the Hubbard case (inset of Fig. 1(c)). Similar tiny gap exists in other first-principle models like the iron-adsorbed graphene [15], but to the best of our knowledge, it has not been reported in the TB+Hubbard models. The gap and the total energy variation trend are plotted in Fig. 1(d). We see that when $N_y$ is larger than about 27, the gap becomes almost negligible (less than 1.0 meV), especially at a finite temperature with the thermal fluctuations. So the zSiNR system transits from an insulator to a metal. From later discussion we see there is a topological transition when Ny exceeds 26. In this transition the total energy decreases smoothly, so it is a first-order phase transition.

Similarly, we tune up the SOI values with a fixed ribbon width. We do this comparison because there exist a lot of 2D materials with different intrinsic SOIs [6, 36]. In this case the ribbons width ($N_y$) is set as 10. From Fig. 1(e)-(g) we see that when the SOI increases, the two edge bands become close to each other until they get almost contacted (but still with a very tiny gap, as shown in the inset of Fig. 1(g)). This trend is also plotted in Fig. 1(h). We will see that after the critical SOI value (about 0.16 eV, as denoted by 'A' in Fig. 1(h)) the gap of zSiNR becomes negligible, corresponding to a band insulator-TI transition with the change of the Chern number. However, the total energy decreases smoothly through this first-order phase transition.

We know that the band gap in zSiNR/zGNR result from the flat bands, which reflect the Coulomb repulsion of the edge electrons [28, 32]. In a very wide zGNR, the band gap becomes small [23, 24], which means the Coulomb interaction is relatively weak. And in the bands of Fig. 1, SOI makes an asymmetric dip/peak in the HOMO/LUMO bands. Based on these two facts, it is seen that in case of a wide ribbon or a large SOI, the Coulomb interaction can not compete with the SOI and the energy band gap becomes almost closed.

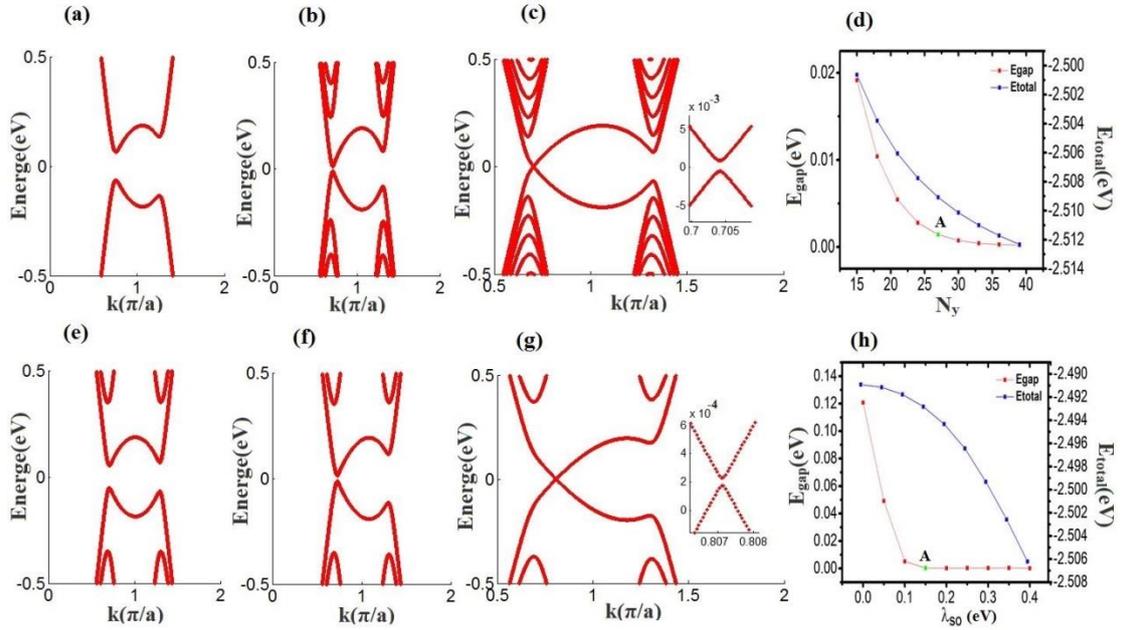

**Figure 1.** Band structures and total energy evolutions with increasing the ribbon width ((a)-(d)) and the SOI ((e)-(h)) in zSiNR. (a)-(c) The band structures of zSiNR ($\lambda_{so}$=0.1 eV) with the ribbon widths ($N_y$) of 5, 15 and 27. (d) The energy gap (left y-axis) and the total energy (right y-axis) dependence on the ribbon width. (e)-(g) The band structures of the zSiNR ($N_y$=10) with the SOI of 0.01eV, 0.07eV and 0.15eV. (h) The energy gap (left y-axis) and the total energy (right y-axis) dependence on the SOI. In (d) and (h) the symbol 'A' denotes the topological phase transition point. The insets in (c) and (g) are the zoomed gaps.

During this insulator-metal transition, we also investigate the band topology of these narrow zSiNRs. The standard QSH and QVH effects in 2D materials come from the non-Hubbard TB model [11, 13]. Here we study these Coulomb-interaction systems with the Hubbard model.

The total Chern number and the spin Chern number are the two characteristic integers to describe the bands' topological properties [8]. To calculate these Chern numbers, the Berry curvature ($\Omega$) is to be evaluated. For the zSiNR, we have to calculate 2D band structures [8, 15]. Similar to the first-principles band calculation for the 1D nanoribbons by the software such as VASP, we design a 2D system, which has the same unit cells of the zSiNR but extends in the x and y directions (Fig. 2(a)). By the Bloch theorem [37], the Hamiltonian of the zSiNR unit cell is written as

$$\mathbf{H}(k_x, k_y) = \sum_{m=-1}^{1}\sum_{n=-1}^{1} \mathbf{H}_{(0,0),(m,n)} e^{i(k_x ma + k_y nb)} \qquad (8)$$

where $\mathbf{H}_{(0,0),(m,n)}$ are the coupling cell matrices between the central unit cell to all the neighboring (including itself) unit cells in a 3*3 supercell system. For example, $\mathbf{H}_{(0,0),(0,0)}$ is the central unit cell, $\mathbf{H}_{(0,0),(-1,0)}$ and $\mathbf{H}_{(0,0),(1,0)}$ are the coupling matrix between the central unit cell to the central-left and central-right unit cell, respectively.

The coupling energies between/within these unit cells (denoted by $t$) are the same as that in the zSiNR, except that in the y direction (denoted by $t_1$). We may tune down $t_1$ for a weak coupling system, which approaches to the qusi-1D zSiNR; and we may set $t_1$ equals to $t$, which is a standard bulk silicene system. So with this coupling-variable 2D system, we could calculate the 2D bands for a quasi-1D zSiNR and thus evaluate the Berry curvature by Eq. (5).

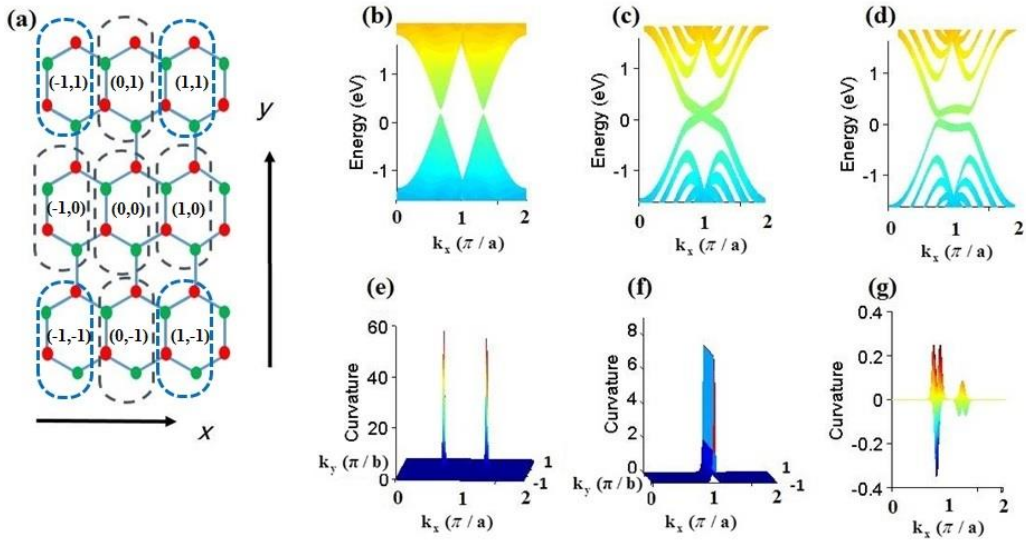

**Figure 2.** (a) Schematic figure of the Berry curvature calculation in the zSiNR. The dashed boxes refer to the central and the neighboring unit cells in the 2D 3*3 supercell system. For simplification, only the zSiNR unit cell with $N_y$=1 is plotted here. The two numbers in the bracket of each unit cell are the (*m,n*) parameters in Eq. (7). (b) and (e) The 2D band structure (side view) and the Berry curvature distribution of the HOMO band (spin up) for the bulk silicene ($t_1$=$t$, $\lambda_{so}$ =0.05 eV). (c) and (f) The 2D band structures (side view) and the Berry curvature distribution of the HOMO band (spin up) for the quasi-1D zSiNR ($t_1$=0.01$t$, $\lambda_{so}$=0.26 eV). (d) and (g) The 2D band structures (side view) and the Berry curvature distribution of the HOMO band (spin up) for the quasi-1D zSiNR ($t_1$=0.01$t$, $\lambda_{so}$ = 0.05 eV). In the band calculations, the TB+Hubbard model is used and the unit cell is similar with that in (a). The geometric and other parameters are set as $N_y$=4, $t$=1.6 eV and U=1.4 eV.

With the SC calculation for the Hubbard system ($t$=1.6 eV, U=1.4 eV, $\lambda_{SO}$ = 0.05eV), we obtain the band structures of a bulk silicene ($t_1$=$t$), as shown in Fig. 2(b). There are small gaps (about $\lambda_{SO}$) around the two Dirac cones (K and K' points). Although the Coulomb interaction is involved, the electron density of each atom equals to 0.5. The Berry curvature ($\Omega$) distribution of the spin-up HOMO band in this bulk silicene has two peaks around the Dirac cones (Fig. 2 (e)). The asymmetry of the two peaks results from the Hubbard term. The integrated curvature gives the Chern number as 1 for the spin-up bands. For the spin down bands, the Chern number is -1. The total and spin Chern number are: (C, $C_s$)=(0, 1). Thus the system belongs to the QSH state as obtained by the non-Hubbard TB model [8].

When tuning down $t_1$ to 0.01$t$, as for a weak-coupling system, we calculate the band and Chern numbers of a quasi-1D zSiNR. Fig. 2(c) and (f) shows the 2D bands and the curvature distribution (for the spin-up HOMO band) with a large SOI value ($\lambda_{so}$ = 0.26 eV). We see that in this case the SOI overcomes the Coulomb interaction and the bands are similar to that of the non-Hubbard TB model (such as Fig. 1(g)). The Chern numbers of two spins are 1 and -1. Thus this system is still a SQH state with the topological numbers (C, $C_s$)=(0, 1).

For the small SOI case ($\lambda_{so}$ = 0.05 eV) of the weak-coupling system, the bands and curvature distribution are shown in Fig. 2 (d) and (g). A band gap is opened due to the Coulomb interaction, which is similar to the band in Fig. 1(e). The $\Omega$ peaks locate at the two extreme points of the bands (k=0.6π/a and 1.4π/a). The calculated Chern numbers of the HOMO bands with two spins are both zero. So with a small $\lambda_{so}$, the Coulomb interaction overcomes SOI and the zSiNR experiences a topological phase transition from the SQH-typed TI to a band insulator. The detailed process for the zSiNR with $N_y$=10 is given in Fig. 1(h). When $\lambda_{so}$ exceeds 0.15 eV (the 'A' point in Fig. 1(h)), the system has a topological transition.

We also check the Chern numbers of the zSiNR for different ribbon widths with $\lambda_{so}$ =0.05eV. We find at $N_y$=26 (the 'A' point in Fig. 1(d)), its topological property transits from the common insulator to the SQH-typed TI.

**3.2 Magnetic phase transition of zSiNR in the exchange fields**

In the previous part we have demonstrated that the Coulomb interaction may overcome the SOI for the narrow ribbons with a small SOI. In this part we focus on these Hubbard-dominated systems and study the magnetic phase transition in the exchange field.

The magnetic exchange field is involved in the zSiNR by the proximity coupling of the ferromagnetic atoms such as Fe and Ni in the substrate[17]. Due to the buckled structure, the exchange field strengths of the neighboring Si atoms are different. So the AF exchange field ($M_{AF}$) or the staggered exchange field is introduced to modify the magnetic strength. Here we firstly omit this difference and only consider the FM exchange field ($M_{FM}$). In the following calculations we use $\lambda_{so}$ value of 3.9 meV for the zSiNR except the special case for a larger SOI value.

Figure 3 shows the band diagrams of zSiNR ($N_y$=3) under different FM exchange fields at zero temperature. In the absence of FM exchange field, due to the Coulomb repulsion between the electrons, the system exhibits the AF state as a ground state (Fig. 3(a)). With the increase of the FM exchange field, the degenerate bands of the two spins are gradually separated in the upper and lower directions (Fig. 3 (b)). Near the critical value ($M_{FM}$=0.125eV), there are some band contacting and merging behaviors of the HOMO and LUMO bands. We find that the spin-up HOMO band (red) and the spin-down LUMO band (blue) are almost close to the Fermi level (E=0) at the critical $M_{FM}$ (Fig.3 (c)). After this critical value, these two bands approach and intersect, the spin-up band enters the conduction band region, and the spin-down band enters the valence band region. After the self-consistent iteration, in order to obtain the minimum energy, the two bands will further rise and fall until they merge with the same spin bands in their respective conduction (valence) band region (Fig.1 (d)). At this time, the number of spin-up bands near k=1.0π/a below the Fermi level is larger, so the spin-down electrons is more than the spin-up electron and the system becomes ferromagnetic.

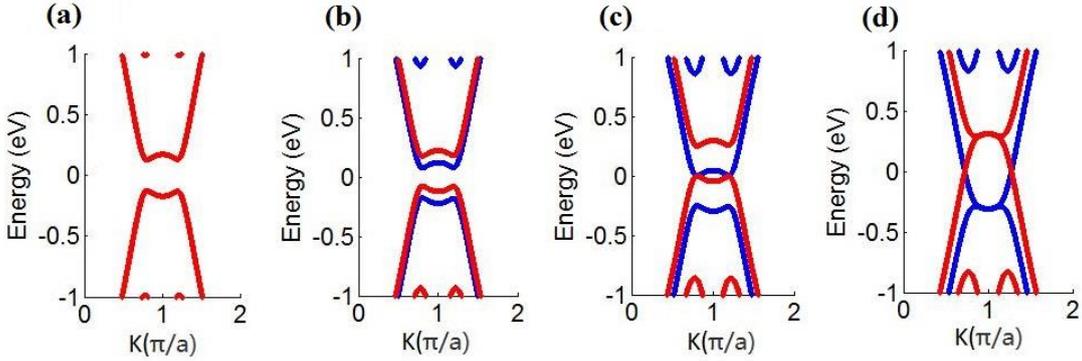

**Figure 3.** Energy band structures of the zSiNR ($N_y$=3) with different FM exchange fields in an initial state of AF order. The $M_{FM}$ increases in (a)-(d) with the values of 0 eV, 0.05 eV, 0.125 eV, and 0.130 eV. The red and blue bands correspond to the spin up and spin down components; $a$ is the lattice constant in the unit cell of band calculations.

The energy band gap ($E_{gap}$) is plotted along with the change of the FM exchange field, as shown in Fig. 4 (a) (blue, T=0K). We see that $E_{gap}$ of the two spins are always degenerate and remains the same before $M_{FM}$=0.125eV. After this value, the gap suddenly drops to zero in a phase transition to FM state. In the meanwhile, at zero temperature we calculate the total energy ($E_{total}$) of the system under different exchange fields (averaged in the Brillouin zone and divided by the number of atoms in the unit cell), such shown in Fig.4 (b) (blue). We see that the energy is a constant before the transition point. After the transition point, the energy curve has a sudden decrease and becomes a tilted straight line. This linearly change of the total energy may be explained by the magnetization energy in the FM state: $E_M = \sum_i M_{FM}(<n_i^\uparrow> - <n_i^\downarrow>) = M_{FM}(N_{occ}^\uparrow - N_{occ}^\downarrow)$. From Fig. 4(c) we see that the number difference of the spin-down and spin-up electrons $(N_{occ}^\uparrow - N_{occ}^\downarrow)$ is a constant, so $E_M$ varies linearly with the FM field. We also see the system is more stable in a larger FM exchange field.

Now we give a quantitative demonstration for the constant energy before the phase transition. With the FM exchange field, the eigen-energy of the spin-up and spin-down electron may be obtained from the following formula

$$(\mathbf{H}^\uparrow + \mathbf{I}_M)\boldsymbol{\varphi}_1 = E_1 \boldsymbol{\varphi}_1$$
$$(\mathbf{H}^\downarrow - \mathbf{I}_M)\boldsymbol{\varphi}_2 = E_2 \boldsymbol{\varphi}_2$$

where $\mathbf{I}_M$ is the Hamiltonian of the FM exchange field, $\mathbf{I}_M = M_{FM}\mathbf{I}$, $\mathbf{I}$ is the unit matrix. $\boldsymbol{\varphi}_1$ ($\boldsymbol{\varphi}_2$) are the eigenvectors of electron with two spins from the SC calculation. Thus we see that

$$E_1 + E_2 = \boldsymbol{\varphi}_1^\dagger (\mathbf{H}^\uparrow + \mathbf{I}_M)\boldsymbol{\varphi}_1 + \boldsymbol{\varphi}_2^\dagger (\mathbf{H}^\uparrow - \mathbf{I}_M)\boldsymbol{\varphi}_2 = \boldsymbol{\varphi}_1^\dagger \mathbf{H}^\uparrow \boldsymbol{\varphi}_1 + \boldsymbol{\varphi}_2^\dagger \mathbf{H}^\uparrow \boldsymbol{\varphi}_2.$$

With averaging in the k-space and summing for all the occupied electrons, the total energy is obtained, which is still independent of $M_{FM}$.

For the process near the phase transition point, we give a detailed analyze: before the two bands (the aforementioned LUMO and HOMO bands) get contact, the numbers of valence orbitals with two spins are equal. Thus the occupied electrons of two spins are the same, as shown in Fig.4(c). After the critical point, partial of the HOMO bands (spin-down) go into the valence band region, whose orbitals are occupied by spin-down electrons, while partial of the LUMO bands (spin-up) enter the conduction band region, whose orbitals are empty for the spin-up electrons. So the number of spin-down electron becomes more than the spin-up electron and the system exhibit a weak FM property. Because the final steady state has a local minimum of the total energy [28], in the iteration calculation process the intersected bands will rise and fall until they merge with their respective 'same-spin' counterpart in the conduction and valence band regions (Fig. 3(d)). Fig. 4(d) shows the energy evolution in this iteration. We see that the total energy rises and reaches a maximum value (at step 5). Before and in this iteration step the system exhibit an AF property (see the first inset). After this step, the total energy drops and the system exhibit a FM property (see the second inset).

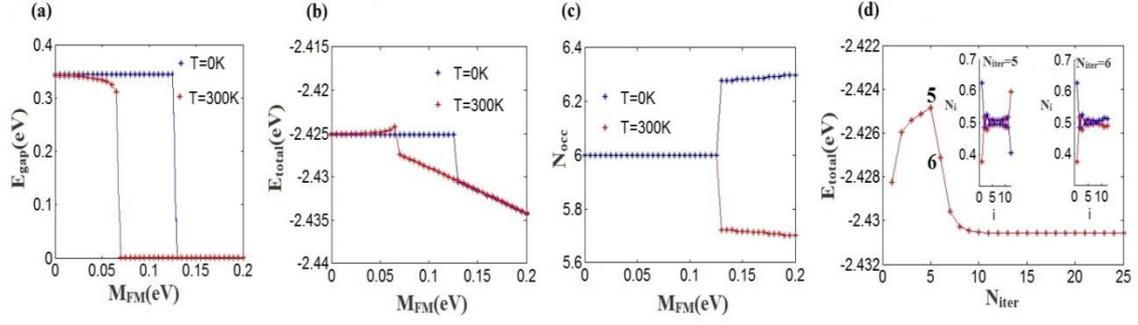

**Figure 4.** AF-FM phase transition analysis for the zSiNR ($N_y$=3) with a FM exchange field. (a) The gap dependence on the FM exchange field at the zero (T=0K, blue) temperature and the room temperature (T=300K, red). (b) The total energy dependence on the FM exchange field at the zero (blue) and the room temperature (T=300K, red). (c) The total occupation number of the electrons (red: spin up; blue: spin down) in zSiNR with different FM exchange fields at T=0K. (d) The total energy variations of zSiNR in each iterations steps of the SC band calculation ($M_{FM}$=0.13 eV and T=0 K). The two insets represent the electron density distributions (red for spin-up electron and blue for spin-down electron) of zSiNR at the iteration step 5 (the AF state) and step 6 (the FM state).

We also investigate this AF-FM phase transition at room temperature (T=300K). Similarly, around some critical FM exchange field, the AF state of zSiNR changes into FM state. We plot the curves of $E_{gap}$ and $E_{total}$ in Fig. 4 (a) and (b) (red curves). The phase transition point (about 0.062eV) at room temperature is much lower than that of the zero temperature. So the increase of temperature is beneficial to this magnetic phase transition. This can be explained from the electron density calculation formula at finite temperature (Eq. (4)). When T>0K, the bands with the energy above the Fermi level also have a small occupation probability. So according to the previous analysis at the zero temperature, before the two special bands meet the Fermi level, the spin-down electrons may become smaller than the spin-up counterpart. Then similar to the Fig. 4(d), the total energy reaches to the maximum point in the iteration and then reaches the local minimum valley. After the SC calculation, the AF-FM transition occurs with a smaller critical $M_{FM}$ value.

With the same approach, we investigate the FM-AF transition for the FM-ordered zSiNR with the AF exchange field. From Fig. 5 (a) and (b) we see with increasing $M_{AF}$, the two spin bands near k=π/a get to split to form a gap. The split of these bands gradually becomes large until a critical value is reached (Fig. 5(c)). After this critical value, the spin-up band (band a in Fig. 5(c)) suddenly rise to merge with the spin-down band (band b in Fig. 5(c)) in the conduction band region to form the AF degenerate bands (Fig. 5(d)). The similar merge occurs for the band c and d in Fig. 5(c). We see this is a reverse process of the AF-FM transition.

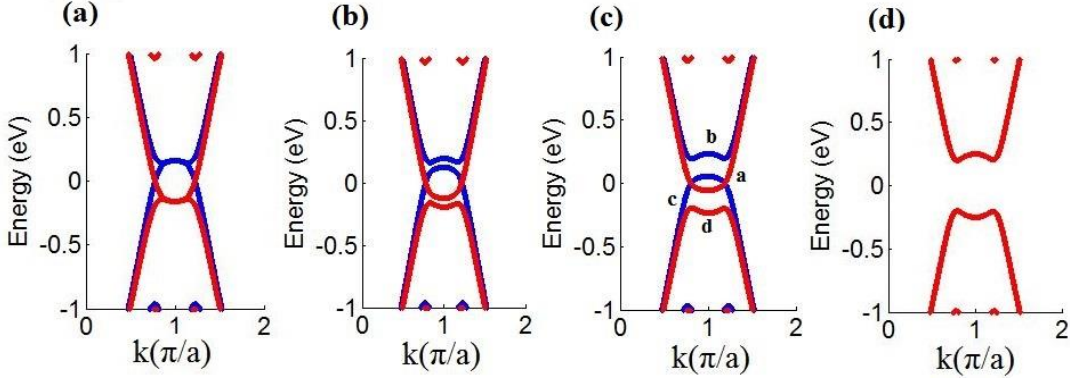

**Figure 5.** Energy band structures of the zSiNR ($N_y$=3) with different AF exchange fields in an initial state of FM order. The $M_{AF}$ increases in (a)-(d) with the values of 0 eV, 0.010 eV, 0.045 eV and 0.050 eV. The red and blue bands correspond to the spin-up and spin-down components.

We also plot the energy gaps of the two spin bands of this zSiNR with the change of the external AF exchange field, as shown in Fig. 6 (a) (blue curve, T=0K). We see that before $M_{AF}$=0.045eV, $E_{gap}$ of the two spins are the same, and linearly increase with $M_{AF}$. After the phase transition point, the FM state changes to the AF state. The energy gap has an abrupt rise, and then the gap increases linearly. Similarly, the total energy ($E_{total}$) curve of the system at zero temperature shows a sudden drop at the phase transition point before a slow decrease (Fig. 6(b)). After this transition, the system becomes the AF state, and the number of the spin-up electron becomes equal to the number of the spin-down electron (Fig. 6(c)). The total system energy decreases linearly with the AF exchange field. We see that the larger of the AF exchange field, the more stable of the system is.

The FM-AF transition of zSiNR at the room temperature is also investigated with the Hubbard model. The energy gap and total energy dependence on the AF exchange field are drawn in Fig. 6 (a) and (b) (red curve). We see the similar behaviors as in the zero temperature case, except that the transition point of $M_{AF}$ (0.025eV) becomes smaller. The reason is similar with the case of the AF-FM transition at the room temperature (see the previous statement in Fig. 4).

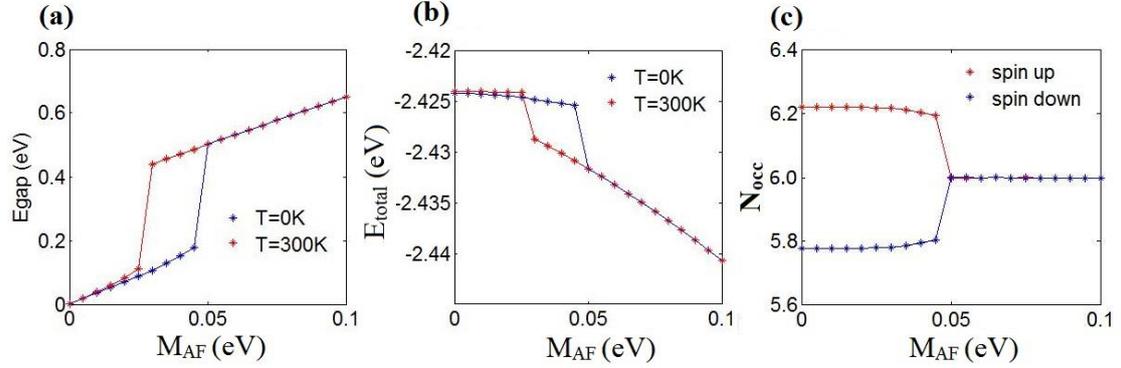

**Figure 6.** FM-AF phase transition analysis for the zSiNR ($N_y=3$) with an AF exchange field. (a) The gap dependence on the AF exchange field at the zero (T=0K, blue) temperature and the room temperature (T=300K, red). (b) The total energy dependence on the AF exchange field at the zero (blue) and the room temperature (red). (c) The total occupation number of the electrons (red: spin up; blue: spin down) in zSiNR with different AF exchange fields at T=0 K.

**3.3 Magnetic/topological phase transitions in zSiNR with electric fields**

Besides the phase transition induced by the FM and AF exchange field, we also investigate the phase transition by the electric field. The electric field is applied perpendicular to the ribbon plane with the strength of $E_z$. The corresponding Hamiltonian is given in Eq. (1).

From Fig.7 (a) we see that at the zero temperature, the degenerate AF bands of zSiNR are split in the up and down directions under the small electric potential ($E_z l$=0.1eV). This split continues until the two spin-down bands get contacted (Fig. 7(b)). From Fig. 7 (d)-(e) we see the spin-up electron densities almost do not change during this time, because the spin-up energy bands (red curves in Fig. 7(a)-(b)) and the corresponding eigenvectors are almost unchanged. As the electric field increases, the spin-down electron densities on the two edge atoms get more close to 0.5. Fig. 5 (c) and (f) show that when the electric potential energy is 0.19eV, the spin-up and spin-down bands coincide with a smaller gap than that of the AF state. Also the electron densities of two spins merge when $E_z l$ is 0.19eV. In this case the AF state is destroyed by the external electric field and the system becomes nonmagnetic.

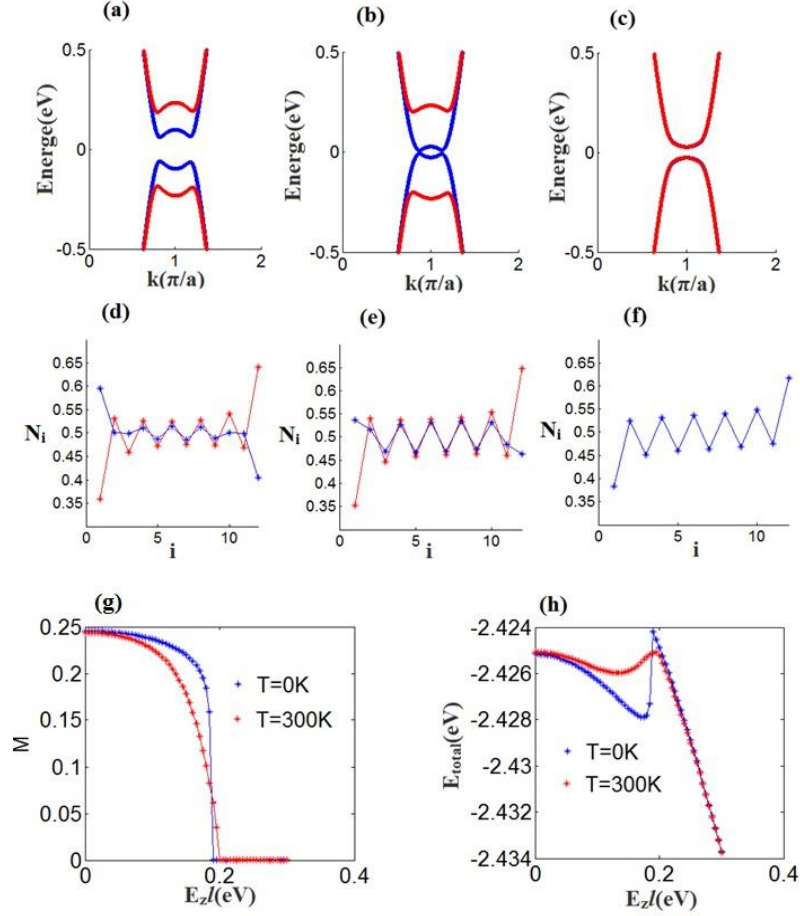

**Figure 7.** Band structures and electron density distributions of zSiNR ($N_y=3$) under different electric fields with an initial AF spin configuration (T=0K). (a) and (d) $E_zl$= 0.10eV; (b) and (e) $E_zl$=0.18eV; (c) and (f) $E_zl$= 0.19eV. (g) and (h) Spin polarization and total energy of zSiNR (Ny=3) dependence on the external electric field. The blue and red curves correspond to the zero temperature and the room temperature.

We also calculated the spin polarization (M) and total energy with different electric field in this phase transition process (Fig. 7(g)-(h)). M is defined as the average electron-density difference of two spins on the two edge atoms. From these figures it is seen that at zero temperature, with the increase of $E_z$, the spin polarization curve (blue) decreases gradually and then suddenly drops at $E_zl$=0.19eV, as the AF state is destroyed. After that, the density remains zero. In the transition region, the red M curve (at room temperature) decreases more slowly than that the blue curve (at zero temperature). And the critical point of $E_z$ is smaller at room temperature ($E_zl$=0.20eV for at the room temperature). From this we see that a finite temperature will make the AF state more vulnerable to destruction. Similarly, at the room temperature, with the increase of the electric potential energy the total energy becomes decreased, and in the vicinity of the critical point where the AF state is destroyed, $E_{total}$ gradually rises, and then decrease linearly with the increase of $E_z$. But at zero temperature, there is very steep rise of $E_{total}$ near the critical point of the AF-state destruction.

At last we explore the electric effect on the zSiNR system with a larger SOI ($\lambda_{so}$=50 meV) and wider ribbon width ($N_y$=5). As we stated in Sec. IIIA, in this case the SOI can compete with the Coulomb interaction. Figure 8 (a) and 8(b) shows the band structures at the $E_zl$=0.15 eV and 0.20 eV respectively. Similar to Fig. 7(a), in a very small electric field, the spin-up HOMO and LUMO bands approaches to each other until they get contacted ($ka \approx 0.6\pi$ in Fig. 8(a)). With a larger electric field, the two contacted bands get separated again (Fig. 8(b)). The band gap and spin polarization dependence on the electric field are also plotted in Fig. 8 (c). From this figure we see that the band gap of the system decreases to about zero with $E_zl$=0.14eV and then the gap opens again with $E_zl$=0.19eV. After the gap opens, there is a magnetic phase transition in zSiNR, as we see that the spin polarization curve becomes zero and the AF-ordered system becomes non-magnetic.

The corresponding Berry curvatures for the spin-up HOMO bands with $E_zl$=0.15eV and 0.20 eV are calculated and drawn in Fig. 8 (d) and (e) respectively. In a comparison with Fig. 8(a) we see that in Fig. 8(d) there is a sharp peak around the contacting point of two spin-up edge bands. The integrated Chern number is 1.0 for the spin-down band and 0 for the spin-up band. We see that in this case the zSiNR has the topological number (C, $C_s$)=(1, $\frac{1}{2}$). It belongs to the spin quantum anomalous Hall (SQAH) TI [8]. Similar results can be found in Z.M. Yu and Y.G. Yao's work with the external Rashba and exchange field [38]. The edge states for the spin-down band below the Fermi level are demonstrated in the inset of Fig. 8 (a). We see that the two opposite-propagating waves localized near the two ribbons edges. While for $E_zl$=0.20eV, the Chern number of the spin-up and spin-down bands are all zero (Fig. 8(e)).

With this quasi-1D Berry curvature calculation ($t_1$=0.001$t$), we evaluate the Chern numbers with difference $E_z$ values, as shown in Fig. 8(f). From this figure we see that besides the magnetic phase transition point ($E_zl$=0.19eV), there exists a topological phase transition point ($E_zl$=0.14eV), where the Chern number changes from 0 to 1. So in this Hubbard-SOI-balanced system, with the perpendicular electric field, there exist two types phase transitions: the topological phase transition and the magnetic phase transition. In a strong electric field, both of the SQAH-typed TI and magnetic phases are destroyed and the system returns to the common non-magnetic band insulator.

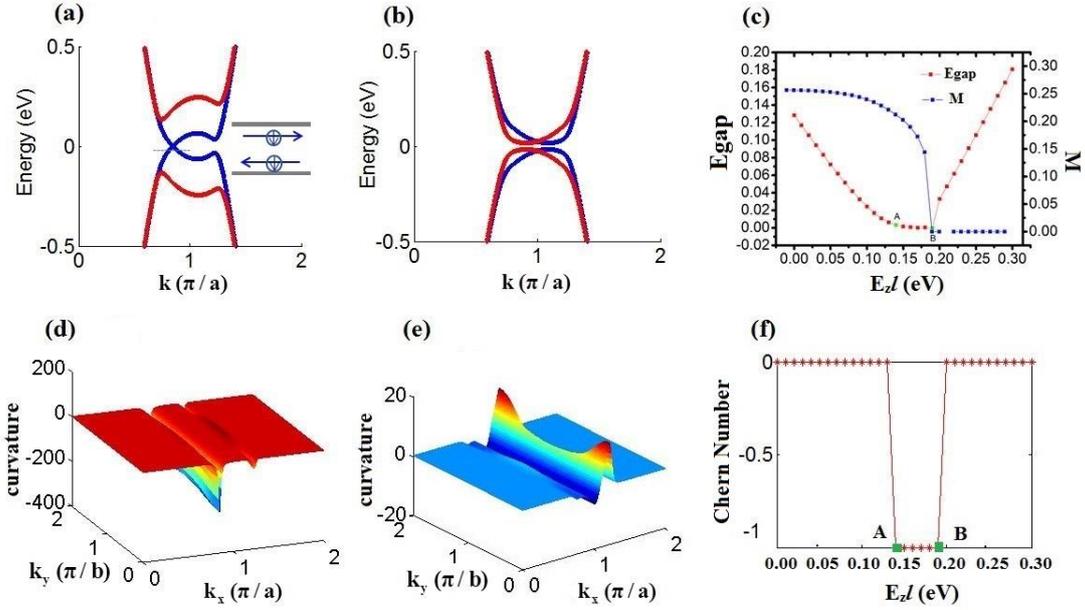

**Figure 8.** Band structures and the topological properties for the Hubbard-SOI-balanced system with an electric field. The system is zSiNR with $N_y$=5 and $\lambda_{so}$=50 meV. (a) and (b) The energy bands with $E_zl$=0.15 eV and $E_zl$=0.20 eV; The edge states of spin-down electrons below the Fermi level are plotted in the inset of (a); (c) The energy gap and spin polarization dependence on the electric field; (d) and (e) The Berry phase distribution of zSiNR with $E_zl$=0.15 eV and $E_zl$=0.20 eV ($t_1$=0.01$t$); (f) The Chern number variations on the electric potential on the zSiNR.

## 4. Conclusions

In this paper we have used the TB+Hubbard model to systematically investigate the topological and magnetic phase transitions in the zSiNR systems. We have found although in the narrow zSiNRs the Coulomb interaction from the Hubbard term results in a gap, in very wide nanoribbons or with a large-enough SOI, the Coulomb repulsion from the edge electrons is relatively weaker than the SOI. So in this case the band structure of the zSiNR systems will transit from the gapped insulator to a quasi-gapless metal with non-trival topological properties of a QSH state.

In the narrow zSiNRs, we have studied the magnetic phase transition with the external exchange field and electric field. Under the FM exchange field, the ground state of AF-ordered zSiNR will transit to the FM state with an abrupt energy drop after some critical FM value. After this transition, the spin-degenerate bands become spin-separated bands. We have used the principle of the minimum total energy to analyze the details of this transition process. At room temperature, we have found a small transition FM field which means the finite temperature always benefits the phase transition. Similarly, we have analyzed the FM-AF transition of zSiNR under an AF exchange field at the zero and room temperature.

We also have found a magnetic phase transition of zSiNR under a perpendicular electric field. After some critical $E_z$ value, the original AF state turns into a non-magnetic state. At zero temperature, the spin polarization (M) suddenly drops to zero across this value, while at the room temperature, M gradually decreases to zero at a higher transition $E_z$ value.

At last, we have investigated the electric-induced transition in the zSiNR system with a larger SOI value. Unlike the Hubbard-dominated system before, we have observed two types of phase transitions: at a lower critical $E_z$ value, the system transits from a Hubbard insulator to a SQAH state; at a higher critical $E_z$ value, the system transits from the SQAH state to a non-magnetic band insulator.

So with the Hubbard term, the combination of the Coulomb interaction and the SOI results in abundant topological and magnetic phases in the zSiNR systems. We believe this phase transition study for zSiNR can benefit not only a better understanding of their physics behaviors, but also a perspective design for the novel spintronics devices.

**Acknowledgements**

The authors thank Prof. Xin Li and Prof. Dezhuan Han in the college of physics, Chongqing University for their helpful discussions on the phase transitions and the Berry phase properties. We thank Prof. Rui Wang for his kind help in the computer services. Financial support from the starting foundation of Chongqing University (Grants No. 0233001104429) and NSFC (11647307) are also gratefully acknowledged.